\documentclass[notitlepage,aps,pra,amsmath,amssymb,superscriptaddress,showpacs]{revtex4-1}

\usepackage{graphicx}
\usepackage{dcolumn}
\usepackage{bm}
\usepackage{hyperref}

\def\be{\begin{eqnarray}}   
\def\ee{\end{eqnarray}}

\begin{document}

\title[Light-induced anomalous Hall effect in open-quantum systems]{Light-induced anomalous Hall effect in massless Dirac fermion systems
and topological insulators with dissipation}

\author{S.~A.~Sato}
\email{ssato@ccs.tsukuba.ac.jp}
\affiliation 
{Center for Computational Sciences, University of Tsukuba, 1-1-1 Tennodai, Tsukuba, Ibaraki, Japan}
\affiliation 
{Max Planck Institute for the Structure and Dynamics of Matter, Luruper Chaussee 149, 22761 Hamburg, Germany}

\author{P.~Tang}
\affiliation 
{Max Planck Institute for the Structure and Dynamics of Matter, Luruper Chaussee 149, 22761 Hamburg, Germany}

\author{M.~A.~Sentef}
\affiliation 
{Max Planck Institute for the Structure and Dynamics of Matter, Luruper Chaussee 149, 22761 Hamburg, Germany}

\author{U.~De~Giovannini}
\affiliation 
{Max Planck Institute for the Structure and Dynamics of Matter, Luruper Chaussee 149, 22761 Hamburg, Germany}

\author{H.~H\"ubener}
\affiliation 
{Max Planck Institute for the Structure and Dynamics of Matter, Luruper Chaussee 149, 22761 Hamburg, Germany}

\author{A.~Rubio}
\email{angel.rubio@mpsd.mpg.de}
\affiliation 
{Max Planck Institute for the Structure and Dynamics of Matter, Luruper Chaussee 149, 22761 Hamburg, Germany}
\affiliation 
{Center for Computational Quantum Physics (CCQ), Flatiron Institute, 162 Fifth Avenue, New York, NY
10010, USA}

\date{\today}

\begin{abstract}
Employing the quantum Liouville equation with phenomenological dissipation,
we investigate the transport properties of massless and massive Dirac fermion systems
that mimics graphene and topological insulators, respectively.
The massless Dirac fermion system does not show an intrinsic Hall effect, 
but it shows a Hall current under the presence of circularly-polarized
laser fields as a nature of a optically-driven nonequilibrium state.
Based on the microscopic analysis, we find that the light-induced Hall effect mainly 
originates from the imbalance of photocarrier distribution in momentum space
although the emergent Floquet-Berry curvature also has a non-zero contribution.
We further compute the Hall transport property of the massive Dirac fermion system
with an intrinsic Hall effect in order to investigate 
the interplay of the intrinsic topological contribution
and the extrinsic light-induced population contribution.
As a result, we find that the contribution from the photocarrier population imbalance
becomes significant in the strong field regime and it overcomes
the intrinsic contribution.
This finding clearly demonstrates that intrinsic transport properties of materials 
can be overwritten by external driving and may open a way 
to ultrafast optical-control of transport properties of materials.
\end{abstract}

\maketitle


\section{Introduction \label{sec:intro}}

Controlling material properties by external driving is one of the ultimate
goals of modern condensed matter physics.
Light is one of the most important drivers
to realize ultrafast control of material properties \cite{Krausz2014,Basov2017}.
A strong terahertz field can couple with a specific phonon mode in solids
and largely populate the phonon mode.
As a result of the significant phonon excitation, electron-phonon coupling is renormalized,
and a superconducting state is realized in the picosecond time scale \cite{Fausti189,Mitrano2016}.
Furthermore, recently it was theoretically demonstrated that an optically-driven phonon
can induce magnetism in two-dimensional materials \cite{Shin2018}.
An intense infrared laser field can strongly couple also with
electrons in solids and significantly renormalize electronic structures
within a optical cycle. The light-induced modulation of electronic structure in solids has been suggested theoretically as a means to modify effective interactions \cite{mentink_ultrafast_2015,tancogne-dejean_ultrafast_2018,Sentefeaau6969,topp_all-optical_2018,golez_multi-band_2019} and possibly tune between competing phases of strongly correlated electron systems \cite{kennes_light-induced_2018,Ruggenthaler2018,claassen2018universal}. On sub-femtosecond time scales, even subcycle electron dynamics in solids can now be investigated with attosecond experimental techniques towards petahertz optoelectronics
\cite{Schultze1348,Lucchini916,Mashiko2016,Schlaepfer2018}.

In contrast to the subcycle electron dynamics in solids, the cycle-averaged dynamics is 
another important subject and has been intensively studied in a number of 
theoretical works in terms of Floquet theory
\cite{PhysRevB.79.081406,PhysRevB.84.235108,Lindner2011,PhysRevLett.110.026603,Sentef2015,PhysRevB.93.144307,Huebener2017},
where Floquet states $|\Psi^F\rangle$ are stationary solutions of Schr\"odinger equation 
with a time-periodic Hamiltonian, $H(t)=H(t+T)$.
In a seminal work by Oka and Aoki \cite{PhysRevB.79.081406}, it was theoretically
demonstrated that the topology of condensed-matter systems can be controlled by
light via the Floquet engineering.
Furthermore, based on the emergence of topology,
the light-induced anomalous Hall effect in graphene, analogous to the Haldane model \cite{haldane1988model} with broken time-reversal symmetry \cite{PhysRevB.99.235156}, has been predicted.
Motivated by this theoretical prediction, McIver \textit{et al.} recently
observed the light-induced anomalous Hall effect in graphene
under the presence of a circular laser field \cite{2018McIver}.
Soon after, we have theoretically investigated the microscopic origin of
the observed light-induced anomalous Hall effect by the quantum
Liouville equation with phenomenological dissipation \cite{2019Sato}.
As a result, we clarified that the imbalance of photocarrier distribution
of topological Floquet states
in the Brillouin zone predominantly causes
the light-induced anomalous Hall current 
with a small contribution from the emergent Berry curvature of the Floquet states.

The anomalous Hall effect induced by circularly polarized light has been heavily discussed
in various theoretical works
\cite{PhysRevB.79.081406,PhysRevLett.113.266801,PhysRevLett.116.026805,PhysRevB.94.121106,
PhysRevLett.117.087402,PhysRevB.96.041205,PhysRevB.97.035153}
from the point of view of intrinsic and light-induced Berry curverture contributions
in order to explore a possibility of diagnosing and steering topological properties by light.
However, despite the great interest in the subject, population effects with dissipation
have only recently been discussed more seriously in the context of Floquet and topological engineering in open \cite{dehghani_dissipative_2014,seetharam_controlled_2015} and closed \cite{abanin_rigorous_2017,weidinger_floquet_2017,schueler_quench} nonequilibrium many-body systems.
Dissipation and population effects are definitely important in explaining the recent
experiments on the light-induced Hall effect \cite{2018McIver,2019Sato}, and here we will provide an in-depth discussion of them.
In this work, we first discuss appropriate treatment of phenomenological relaxation
for dissipative light-driven condensed matter systems,
taking care of electronic decoherence and thermalization via electron-electron
and electron-phonon scattering.
Then, we investigate the light-induced anomalous Hall effect in a massless
Dirac fermion system as a model of graphene
and elucidate the population contribution and the topological
contribution to the Hall effect.
Finally, we investigate the light-induced anomalous Hall effect in a massive
Dirac fermion system as a simple model describing a topological insulator
in order to explore the interplay of the intrinsic topological
contribution and the extrinsic light-induced contribution to the Hall current.
As a result, we find that the intrinsic contribution can be weakened by strong
optical driving and the extrinsic contribution dominates the properties of
the driven system in the strong field regime.
These findings indicate robustness and generality of the photocarrier population effect
and open a possibility to control of material properties via population-control
on top of state-control by light.

This paper is organized as follows: In Sec.~\ref{sec:method}
we describe the theoretical modeling of massless and massive Dirac fermion
systems with the phenomenological relaxation.
In Sec.~\ref{sec:result} we first assess phenomenological relaxation
constructed with two kinds of physical basis sets: One is the static Bloch state,
and the other is the dynamical Houston state.
Then, we investigate the Hall transport property of massless and massive Dirac fermion 
systems under the presence of circularly polarized laser fields.
Finally, our findings are summarized in Sec.~\ref{sec:summary}.

\section{Method \label{sec:method}}

Here we describe the simple but realistic theoretical modeling of 
massless and massive Dirac fermion
systems. The systems are described by
the following widely used Hamiltonian at each $\vec k$-point,
\be
H_{\vec k} = \hbar v_F \tau_z \sigma_x k_x + \hbar v_F \sigma_y k_y
+ \frac{\Delta}{2}\sigma_z,
\ee
where $\sigma_{u=x,y,z}$ is a Pauli matrix, $v_F$ is the Fermi velocity,
and $\Delta$ is the band gap of the system. If $\Delta$ is set to
zero, the system corresponds to a massless Dirac fermion system. Otherwise, 
the system is a massive Dirac fermion system. 
The chirality of the system is given by $\tau_z=\pm 1$.
Note that the positive and negative chiralities correspond to
the Dirac cone of graphene at $K$ and $K'$ points, respectively.
Thus, one needs to include both contributions in the simulation
to accurately describe the low energy states of graphene.
In this work, we fix the Fermi velocity $v_F$ to $1.12\times 10^6$~m/s,
which corresponds to that of graphene computed by the
\textit{ab-initio} calculation \cite{PhysRevLett.101.226405}.

To describe the light-driven electron dynamics in massive and massless Dirac
fermion systems with dissipation, we employ the following quantum Liuoville equation 
for the reduced single-particle density matrix,
\be
\frac{d}{dt}\rho_{\vec K(t)}(t)=
\frac{\left [ H_{\vec K(t)},\rho_{\vec K(t)}(t) \right ]}{i\hbar}
+\hat D_{\vec K(t)}\rho_{\vec K(t)}(t),
\label{eq:liouville}
\ee
with a time-dependent Hamiltonian $H_{\vec K(t)}$ and 
a phenomenological relaxation operator $\hat D_{\vec K(t)}$.
The time-dependent Hamiltonian is given as
\be
H_{\vec K(t)} = \hbar v_F \tau_z \sigma_x K_x(t) + \hbar v_F \sigma_y K_y(t)
+ \frac{\Delta}{2}\sigma_z,
\ee
where the light-matter coupling is described by the Peierls substitution,
$\vec K(t)=\vec k+e\vec A(t)/\hbar c$,
with a spatially uniform vector potential $\vec A(t)$.

We construct the relaxation operator $\hat D_{\vec K(t)}(t)$ 
based on the relaxation time approximation \cite{PhysRevLett.73.902}.
For this purpose, we first express the reduced density matrix
$\rho_{\vec K(t)}(t)$
on the basis of Houston states \cite{PhysRev.57.184,PhysRevB.33.5494}, 
which are eigenstates of the instantaneous
Hamiltonian $H_{\vec K(t)}$ at each instance;
$H_{\vec K(t)}|u^H_{b \vec k}(t)\rangle=\epsilon_{b\vec K(t)}|u^H_{b \vec k}(t)\rangle$,
where $b$ denotes the band index, valence ($b=v$) or conduction ($b=c$) bands.
The reduced density matrix can be expanded with the Houston states 
at each $\vec k$-point as
\be
\rho_{\vec K(t)}(t) =\sum_{bb'} \rho_{bb',\vec K(t)}
\big|u^H_{b \vec k}(t)\big\rangle \big\langle u^H_{b' \vec k}(t)\big|,
\ee
which can be written in matrix form
with the Houston basis,
\be
\rho_{\vec K(t)}(t):=
\left(
    \begin{array}{cc}
      \rho_{vv,\vec K(t)}(t) & \rho_{vc,\vec K(t)}(t) \\
      \rho_{cv,\vec K(t)}(t) & \rho_{cc,\vec K(t)}(t)
    \end{array}
  \right).
\ee

In the basis of the Houston states, we further construct the relaxation operator 
$\hat D_{\vec K(t)}$ with the phenomenological
relaxation time, $T_1$ and $T_2$, as
\be
\hat D_{\vec K(t)}\rho_{\vec K(t)} :=-
\left(
    \begin{array}{cc}
      \frac{\rho_{vv,\vec K(t)}(t)-\rho^{eq}_{v\vec K(t)}}{T_1} & 
      \frac{\rho_{vc,\vec K(t)}(t)}{T_2} \\
      \frac{\rho_{cv,\vec K(t)}(t)}{T_2} &
      \frac{\rho_{cc,\vec K(t)}(t)-\rho^{eq}_{c\vec K(t)}}{T_1}
    \end{array}
  \right), \nonumber \\
\ee
where $\rho^{eq}_{b\vec K(t)}$ is the Fermi-Dirac distribution,
\be
\rho^{eq}_{b,\vec K(t)} = \frac{1}{e^{\left (\epsilon_{b\vec K(t)}-\mu \right)/k_B T_e}+1}
\ee
with the electron temperature $T_e$ and the chemical potential $\mu_e$.
In this work, we fix the electron temperature to $80$~K and the chemical
potential to zero unless stated otherwise.
Here, $T_1$ denotes the longitudinal relaxation time, which is responsible
for the population decay, while
$T_2$ denotes the transverse relaxation time, which is responsible for decoherence.
According to Ref.~\cite{2019Sato}, the relaxation times, $T_1$ and $T_2$,
are set to $100$~fs and $20$~fs, respectively.
Here, we choose the longitudinal relaxation time $T_1$ according to the electron 
thermalization time scale, while the transverse relaxation time $T_2$ according to
the electron-electron scattering time scale
\cite{PhysRevB.83.153410,Brida2013,PhysRevLett.115.086803}.

Instead of the use of Houston states $|u^H_{v\vec k}(t)\rangle$, 
one may consider to employ static Bloch states,
which are eigenstates of the Hamiltonian,
$H_{\vec k}|u^B_{b \vec k}\rangle=\epsilon_{b\vec k}|u^B_{b \vec k}\rangle$,
to construct the dissipation operator.
However, as will be shown later,
the dissipation operator based on the Bloch states
fails to describe fundamental properties of materials
because the field-induced intraband motion is not taken into account
in the Bloch basis.

Employing the time-evolving reduced density matrix $\rho_{\vec K(t)}(t)$,
one can compute dynamics of a general one-body observable $\hat O$ as
\be
\langle \hat O(t) \rangle = \frac{1}{(2\pi)^2}\int d\vec k
\mathrm{Tr}\left [\hat O \rho_{\vec K(t)}(t) \right ].
\label{eq:expectation-value}
\ee
In this work, we focus on the transport property of massive and massless
Dirac fermion systems. Thus, we compute the time-evolution of
the electric current, employing the following current operator,
\be
\hat J_i(t) = -\frac{c}{\hbar}\frac{\partial H_{\vec K(t)}}{\partial  A_i},
\label{eq:current-operator}
\ee
where $A_i$ is the $i$-th component of the vector potential $\vec A(t)$.

\section{Results \label{sec:result}}

\subsection{Dissipation constructed with Houston and Bloch states}
Here, we elucidate the two kinds of dissipation operators $\hat D$ to
demonstrate that taking the appropriate basis for the density matrix is crucial:
One is constructed with the Houston states, as explained in Sec.~\ref{sec:method}.
The other is constructed with the static Bloch states, which are eigenstates
of the field-free Hamiltonian, instead of using the Houston states.
For the sake of the investigation, we first evaluate the direct current (DC) transport
property of a massless Dirac fermion system, i.e. a model for graphene.

To evaluate the DC transport property, 
we compute the electric current under a static source-drain field.
To smoothly apply the source-drain field $\vec E_{SD}(t)$, 
we employ the following time-profile that includes a switch-on process,
\be
\vec E_{SD}(t) = E_{SD} \vec e_x f\left ( \frac{t}{T_{switch}} \right),
  \label{eq:source-drain-field}
\ee
where the switching function $f(x)$ is defined as
\be
f(x) =
\left\{
  \begin{array}{@{}ll@{}}
    1, & 1<x \\
      3x^2 -2x^3
 , & 0<x\le 1 \\
    0 , & \mathrm{otherwise} 
  \end{array}\right. \nonumber \\
\ee
Here, we set the source-drain direction to the $x$-direction,
the source-drain field strength $E_{SD}$ to 
$10^3$~V/m, and the switching time $T_{switch}$ to $200$~fs.

\begin{figure}[htbp]
  \centering
  \includegraphics[width=0.6\columnwidth]{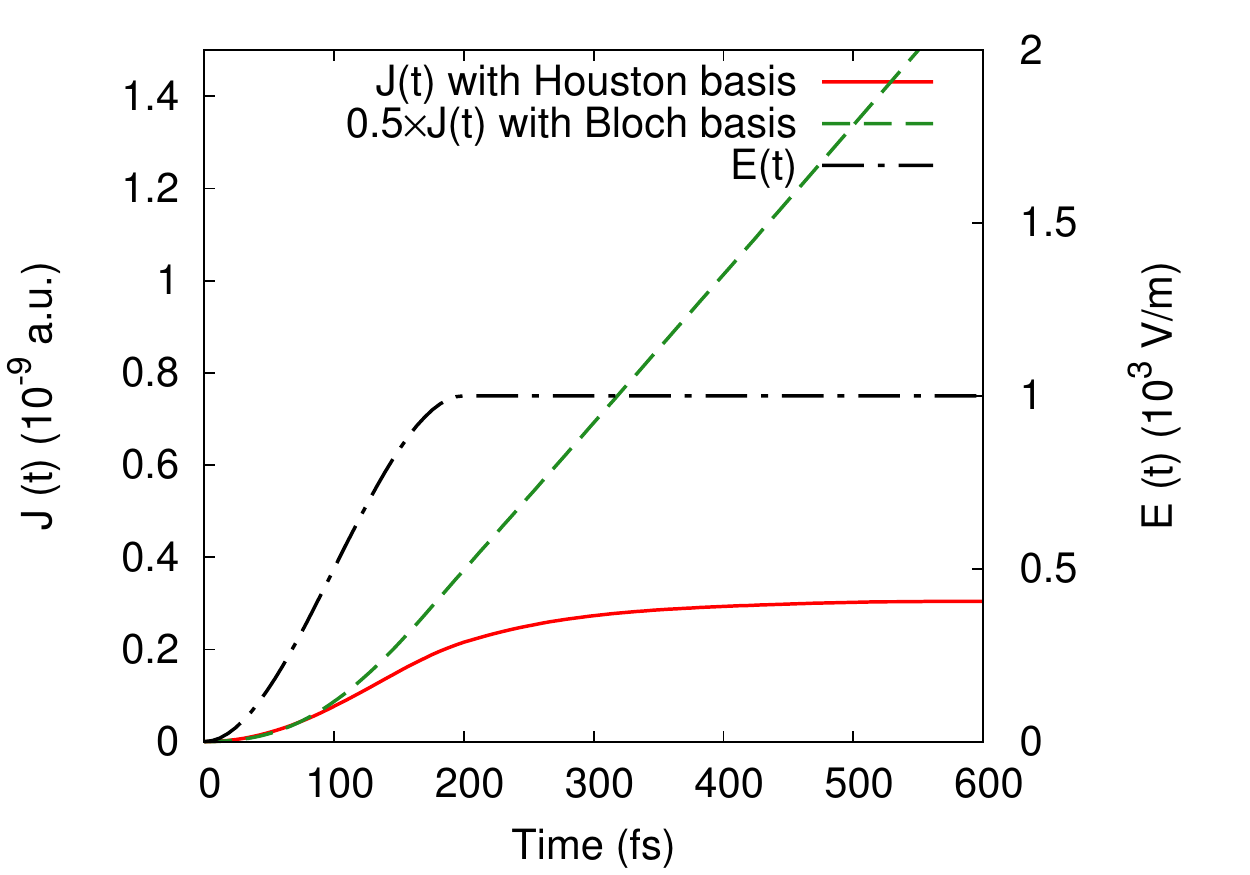}
\caption{\label{fig:dc_current_graphene}
Current as a function of time in a massless Dirac fermion system
under a weak quasi-static electric field.
The red-solid line shows the result with the phenomenological dissipation
constructed with the Houston basis, and the green-dashed line
shows that constructed with the Bloch basis. The black dash-dot line
shows the applied electric field~(right axis).
}
\end{figure}

Figure~\ref{fig:dc_current_graphene} shows the computed current
along the source-drain field direction as a function of time.
The red-solid line shows the result employing
the relaxation operator constructed with the Houston basis,
while the green-dashed line shows that with
the Bloch basis. The time-profile of the applied electric field
is also shown as the black dash-dot line.
One sees that the result using the Houston states~(red-solid)
converges to the constant current on the time scale of the relaxation.
Therefore, the system reaches a nonequilibrium steady state under 
the presence of the source-drain
field, and the saturated current can be described as a function
of the applied field strength, $J(E_{SD})$.
In the weak field limit, the perturbative expansion can be applied,
and one obtains $J(E_{SD})=\sigma E_{SD}$.
Indeed, we numerically confirmed that the current in 
Fig.~\ref{fig:dc_current_graphene} is already in this linear response regime.
Thus, we can confirm that the dissipation operator constructed with the Houston states
can naturally reproduce Ohm's law.
On the other hand, the result using the Bloch states~(green-dashed)
monotonically increases and diverges in the long time propagation limit.
This fact means that the system does not reach a steady state,
demonstrating that the relaxation operator constructed with the Bloch states
induces an artificial error, violating Ohm's law.

Next, we consider the Hall transport property with the dissipation
constructed with the Houston and the Bloch states.
For this purpose, we set the gap $\Delta$ to $100$~meV, and
the chirality $\tau_z$ to $+1$. All the other parameters are the same
as for the above DC transport investigation with the massless Dirac fermion system.
We compute the Hall current along the $x$-axis
under the presence of the slowly-switched-on electric field along the $y$-axis.
Here, we define the instantaneous
Hall conductivity $\sigma_{xy}$ as a ratio of the instantaneous Hall current 
and the strength of the source-drain field $E_{SD}$.

\begin{figure}[htbp]
  \centering
  \includegraphics[width=0.6\columnwidth]{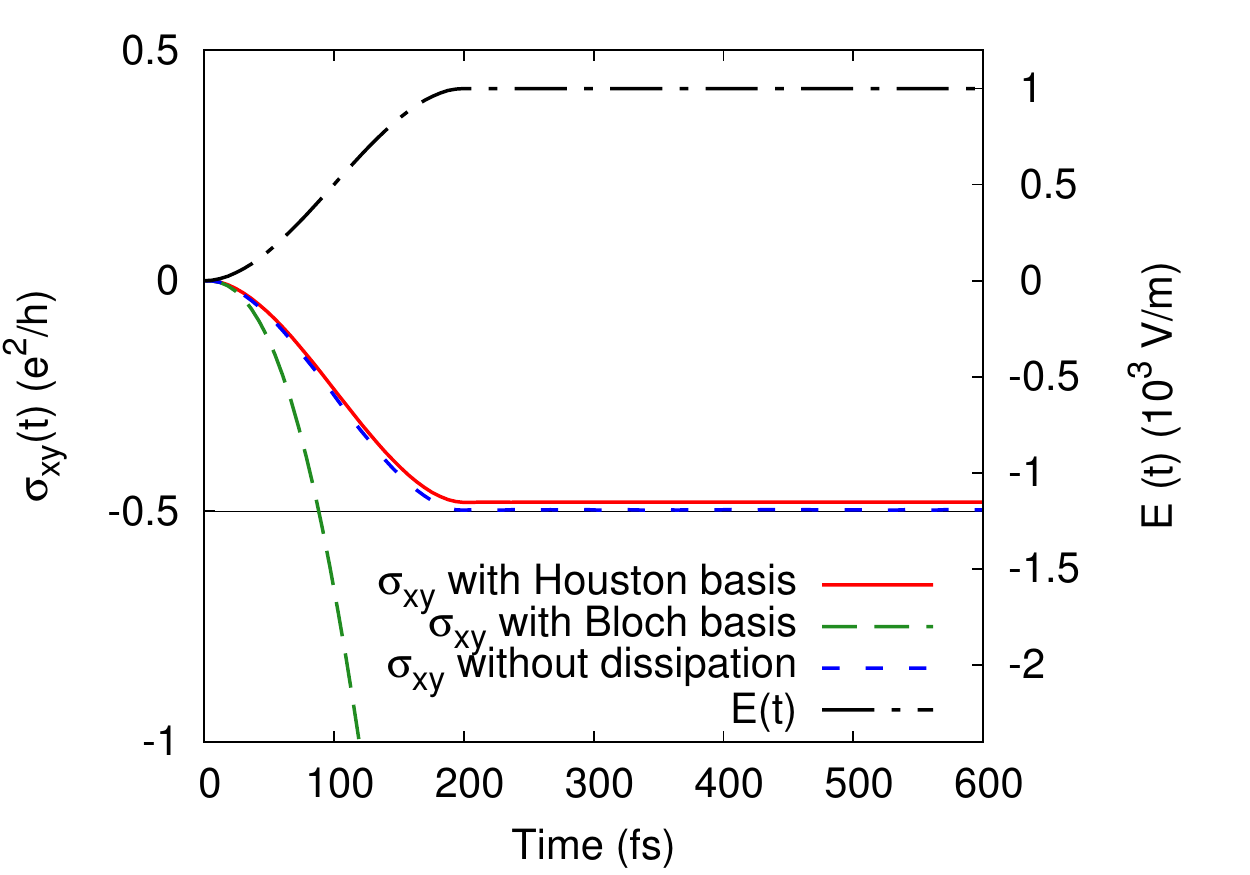}
\caption{\label{fig:hall_current}
Hall conductivity as a function of time in a massive Dirac fermion system
under a weak quasi-static electric field.
The red-solid line shows the result with the phenomenological dissipation
constructed with the Houston basis, the green-dashed line shows that constructed
with the Bloch basis, and the blue-dotted line shows that without any dissipation.
The black dash-dot line shows the applied electric field.
}
\end{figure}

In Fig.~\ref{fig:hall_current}, the red-solid line shows 
the result computed with the dissipation
constructed with the Houston states, while the green-dashed line shows
that with the Bloch states. As a reference, the blue-dotted line
shows the result without any dissipation.
The Hall conductivity without any dissipation converges
to $-e^2/2h $ under the static electric field, showing the half integer
of the quantized conductivity in agreement with the basic theory of
anomalous Hall effect \cite{RevModPhys.82.1959}.
The Hall conductivity computed with relaxation in the Houston basis
fairly reproduces the quantized conductivity,
indicating that relaxation does not have any significant effect
in the regime of adiabatic evolution.
On the other hand, the Hall conductivity computed with relaxation
based on the static Bloch basis significantly deviates from the quantized value,
showing divergent behavior.
Therefore, the relaxation operator constructed with the Bloch states fails
to capture the transport property of a topological insulator.

The failure of the relaxation operator constructed with the static Bloch states can be
understood in terms of the following unphysical excitation mechanism:
Under a static electric field, the vector potential $\vec A(t)$ monotonically
increases and the Hamiltonian $H_{\vec K(t)}$ changes in time.
If the system has a substantial gap, the dynamics has to be adiabatic and the states
have to be well approximated by the eigenstates of the instantaneous Hamiltonian.
However, once the relaxation operator constructed with the static Bloch states is
applied, the operator forces the system to remain in the initial Bloch states,
which are not the eingenstates of the instantaneous Hamiltonian.
Therefore, the relaxation operator using the Bloch states disturbs 
the adiabatic dynamics, and the system is artificially excited.
This unphysical excitation appears as the diverging current
in the DC transport of the massless Dirac fermion system
(in Fig.~\ref{fig:dc_current_graphene})
and in the Hall transport of the massive Dirac fermion system
(in Fig.~\ref{fig:hall_current}).
In contrast, the Houston states can naturally capture this 
adiabatic dynamics since they are defined as instantaneous eigenstates
of the Hamiltonian $H_{\vec K(t)}$.
Therefore, the relaxation operator constructed with the Houston states
does not disturb the adiabatic dynamics under a static electric field,
and it can properly describes the dissipative property of dynamical systems.

Here we examined the properties of the relaxation operator
constructed with the Houston states only for the quasi-static responses as shown
in Fig.~\ref{fig:dc_current_graphene} and Fig.~\ref{fig:hall_current}.
However, importantly, we note that the same description of the relaxation
has been recently examined even for time-dependent fields in both linear and nonlinear
response regimes and has been demonstrated to fairly capture the experimentally observed
features in the context of the energy loss spectroscopy and the high-order
harmonic generation from solids \cite{PhysRevA.97.011401,PhysRevB.99.224301}.
Thus, hereafter, we employ the relaxation operator constructed with
the Houston states in order to investigate the light-induced Hall responses
under the simultaneous presence of static and dynamical fields.

\subsection{Light-induced Hall effect in graphene}

\begin{figure}[htbp]
  \centering
  \includegraphics[width=0.6\columnwidth]{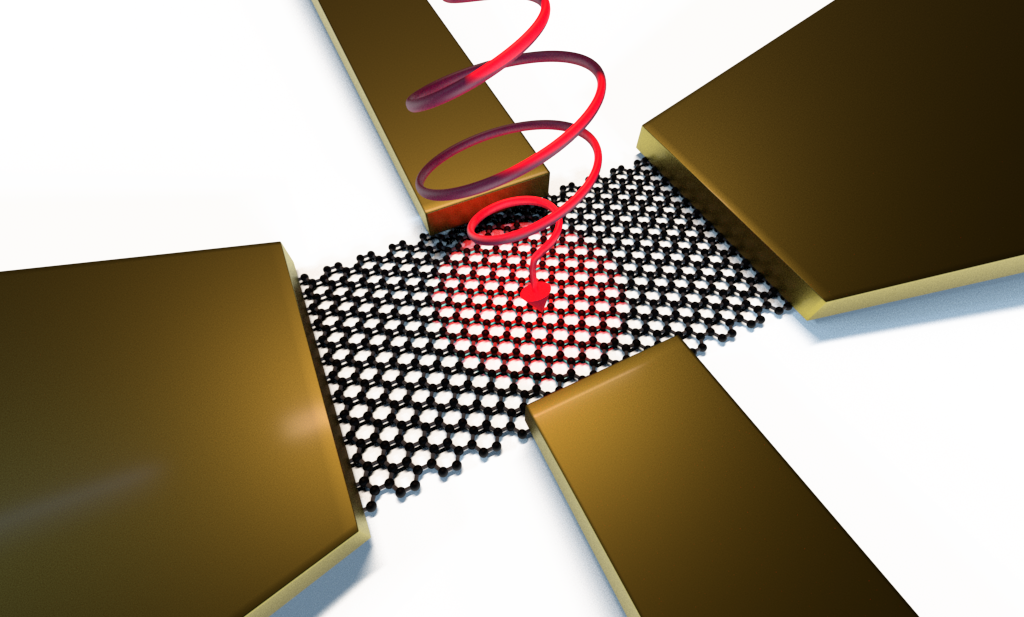}
\caption{\label{fig:fig_trarpes_mod}
Schematic picture of the experimental setup for the light-induced
anomalous Hall effect in graphene.
}
\end{figure}

Next, we investigate the anomalous Hall effect induced
by circularly polarized light in a massless Dirac fermion system
that models the Dirac bands in graphene.
Figure~\ref{fig:fig_trarpes_mod} shows a schematic picture of
our simulation setup that mimics
the experimental setup of Ref.~\cite{2018McIver}. 
We compute the electron dynamics
under the presence of circularly-polarized light and a static
source drain field, and we compute the Hall current, which
is perpendicular to the source-drain direction.

To practically compute the light-induced Hall current,
we employ the following form for a circular laser pulse,
\be
\vec A_{cir}(t)&=&-\frac{cE_{cir}}{\omega_{cir}}
\left [
  \cos \left (\frac{\pi t}{T_{cir}}  \right )
  \right]^2 \nonumber \\
&\times&
\left [
  \vec e_x \sin(\omega_{cir} t)
  + \tau_{cir} \vec e_y \cos(\omega_{cir} t)
  \right ]
\ee
in the domain $-T_{cir}/2<t<T_{cir}/2$ and zero outside.
Here, $\hbar \omega_{cir}$ is the mean photon energy,
$T_{cir}$ is the full duration, $E_{cir}$ is
the maximum of the field strength of a circularly-polarized laser field.
The chirality of the circular laser is given by
$\tau_{cir}=\pm1$.
According to previous work \cite{2019Sato},
we set the full width at half maximum of
the pulse to $1$~ps,
the wavelength of the laser to $6.5$~$\mu$m,
which corresponds to the photon energy of 
$\hbar \omega_{cir}\approx 190$~meV.
For the source-drain field,
we employ the form of Eq.~(\ref{eq:source-drain-field}),
setting $E_{SD}$ to $10^4$~V/m and $T_{SD}$ to $20$~fs.

\begin{figure}[htbp]
  \centering
  \includegraphics[width=0.6\columnwidth]{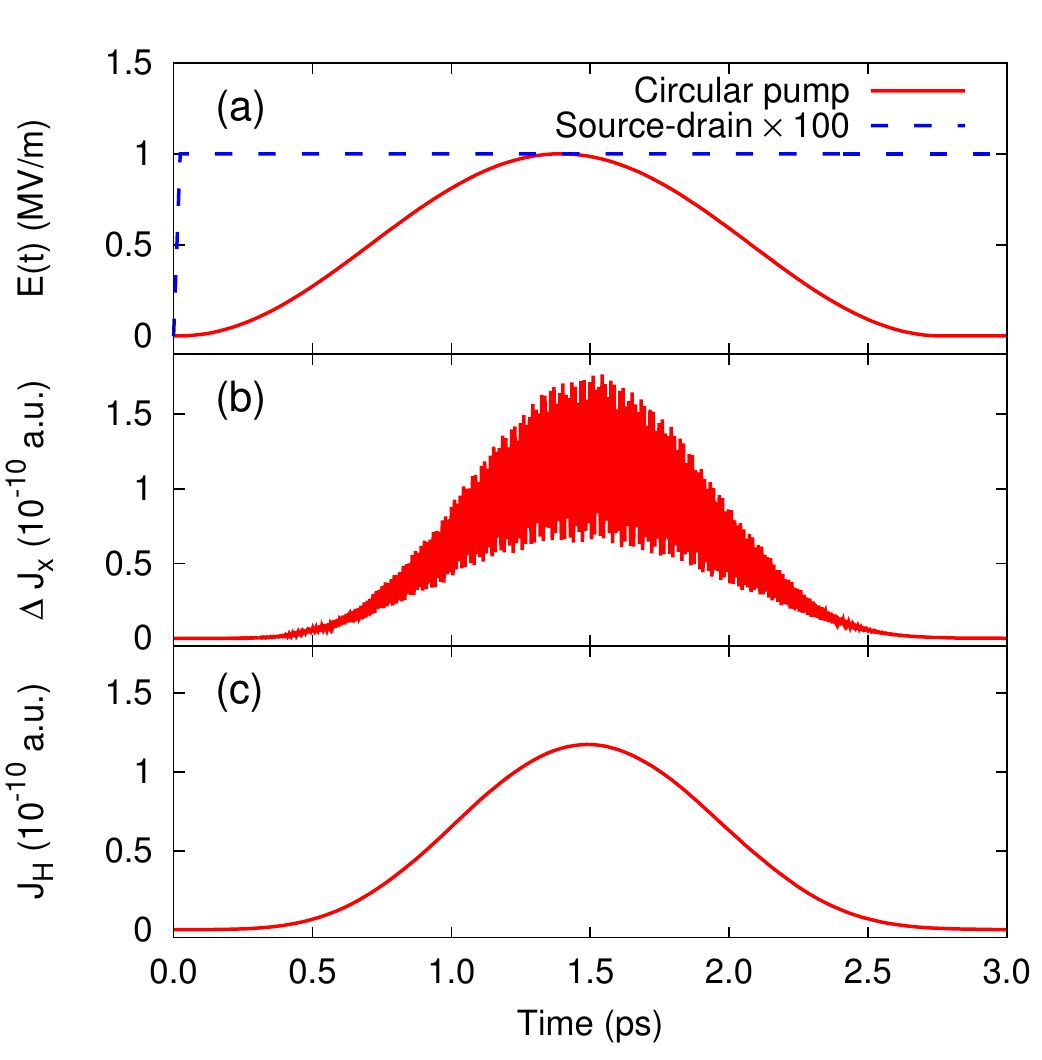}
\caption{\label{fig:hall_dichroism}
Evaluation of the light-induced Hall current in a massless
Dirac fermion system. (a)~The red-solid line shows the field strength 
of the circular laser pulse as a function of time, and
the blue-dotted line shows that of the source-drain field.
Here, the source-drain field (blue-dotted) is scaled by a factor of $100$.
(b)~The difference of the transverse current induced by the positive
and negative circularly polarized laser pulses.
(c)~The extracted Hall current evaluated by Eq.~(\ref{eq:def-hall-current-massless})
with the transverse current in the panel~(b).
}
\end{figure}

To make the direct connection to the experiment \cite{2018McIver},
we evaluate the light-induced Hall current as the difference of the transverse current
induced by positive and negative chirality pump laser fields (pump dichroism).
Figure~\ref{fig:hall_dichroism}~(a) shows the field strength
profile of a circular laser pulse and the source-drain field
as functions of time,
while Figure~\ref{fig:hall_dichroism}~(b)
shows the difference of the transverse current $\Delta J(t)$ induced by
the positive and negative circular laser pulses
under the presence of the source-drain field.
As seen from Fig.~\ref{fig:hall_dichroism}~(b),
the dichroism current $\Delta J(t)$ shows high-frequency components.
These high-frequency components are not relevant for the transport
property because there is no mean charge flow with the time-average.
To extract the DC component of the transverse current,
we define the Hall current as the temporal-average of the dichroism
current as
\be
J_H(t)=\frac{1}{\sqrt{2\pi \sigma^2_W}}\int dt' \Delta J(t')
\exp\left [ \frac{(t-t')^2}{2\sigma^2_W}\right ],
\label{eq:def-hall-current-massless}
\ee
where the width of the temporal average $\sigma_W$ is
set to $100$~fs.

Figure~\ref{fig:hall_dichroism}~(c) shows the computed
Hall current $J_H(t)$ as a function of time.
One sees that the Hall current shows the similar profile to
the applied circular field.
This fact indicates that the light-induced Hall effect in the present conditions
can be characterized with a quasi-steady-state due to the balance of 
the laser excitation and the relaxation at each instance.
Indeed, it was demonstrated that the Hall conductivity evaluated with the peak
Hall current induced by a circular laser pulse
is well described by that evaluated with a steady state under the continuous
circular laser \cite{2019Sato}.

Next, we investigate the light-induced Hall effect by changing 
the field strength of the circular laser pulses.
For this purpose, we define the light-induced Hall conductivity $\sigma_{xy}$
of the massless Dirac fermion system as the ratio of the peak Hall current $J^{peak}_H$
and the source-drain field strength $E_{SD}$,
$\sigma_{xy}=J^{peak}_H/E_{SD}$.

\begin{figure}[htbp]
  \centering
  \includegraphics[width=0.6\columnwidth]{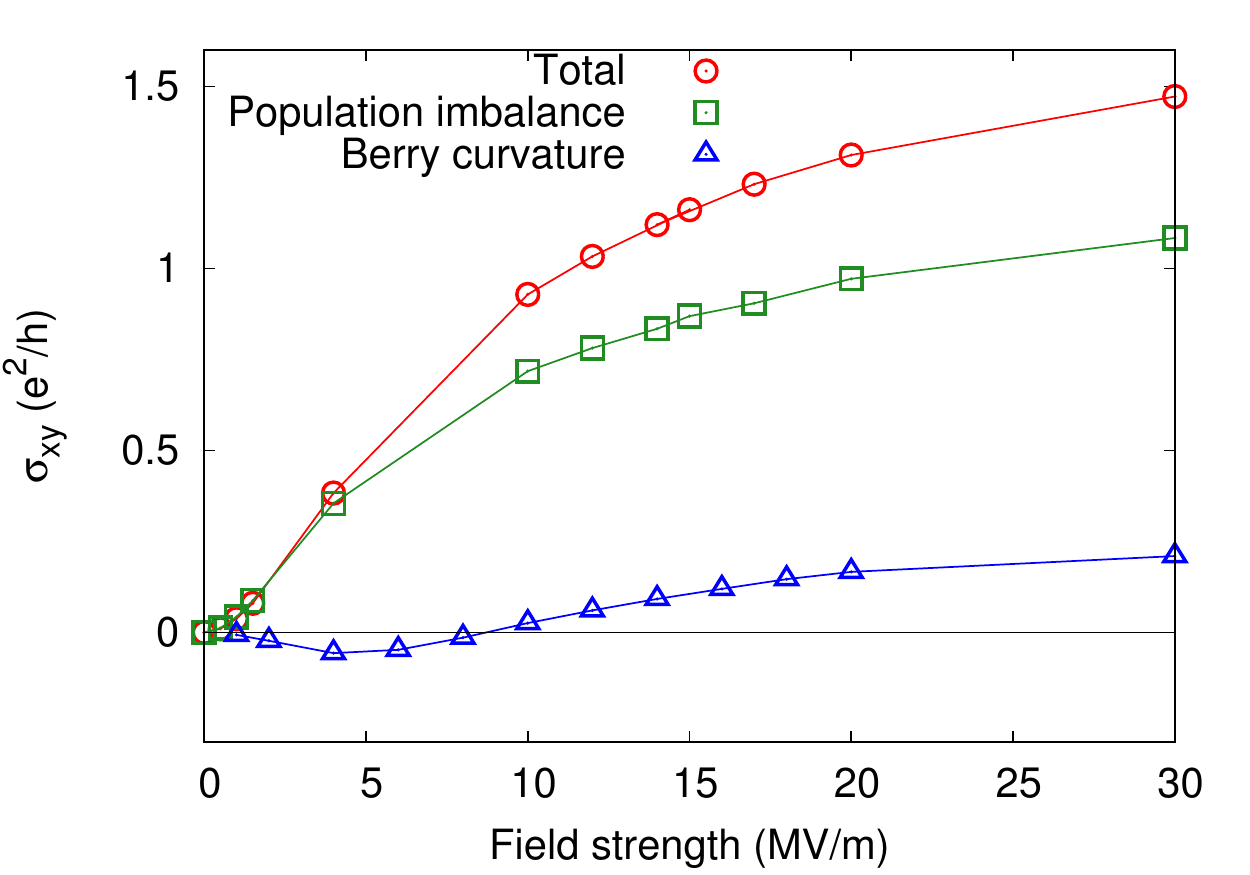}
\caption{\label{fig:intensity_scale}
Light-induced Hall conductivity $\sigma_{xy}$ as a function of the field strength
of the applied circular laser field.
The red circles show the total Hall conductivity computed with
the total current operator, Eq.~(\ref{eq:current-operator}).
The green squares show the contribution from the population imbalance
computed with the intraband current, Eq.~(\ref{eq:intra-current}).
The blue triangles show the contribution from the Berry curvature of
natural orbitals computed with Eq.~(\ref{eq:signa-berry-curvature}).
}
\end{figure}

In Fig.~\ref{fig:intensity_scale}, the red circles show the computed Hall 
conductivity as a function of the applied circular laser fields.
One sees that the Hall conductivity monotonically increases and shows
a saturation behavior in the strong field regime.
In the experiment, a similar saturation has been reported
as the saturated conductivity of $\sigma^{exp}_{xy}=1.8\pm0.4e^2/h$
\cite{2018McIver}.
Therefore, our simulation provides a consistent result with previous experimental
work.

In our previous theoretical work \cite{2019Sato}, 
it was demonstrated that the light-induced Hall effect in graphene
dominantly originates from the population imbalance of 
photocarriers in the momentum space.
Here, we revisit this interpretation from a different angle.
For this purpose, we evaluate the contribution of the momentum-space
population imbalance to the Hall current
by the following intraband current
\be
\vec J_{intra}(t) = \frac{1}{(2\pi)^2}\sum_{b=v,c} \int d\vec k
n_{b\vec K(t)}\frac{\partial \epsilon_{b\vec K(t)}}{\partial \vec k},
\label{eq:intra-current}
\ee
where $\vec K(t)$ is the accelerated Bloch wavevector,
$\vec K(t)=\vec k + e\vec A(t)/\hbar c$, $\epsilon_{b\vec k}$
is the eigen-energy of the Dirac Hamiltonian at $\vec k$.
Here, the instantaneous occupation $n_{b\vec K(t)}(t)$ is computed
by the projection onto the Houston state as
\be
n_{b\vec K(t)}(t)=\mathrm{Tr}\left[ \rho_{K(t)}(t) 
\big|u^H_{b\vec k}(t) \big\rangle
\big\langle u^H_{b\vec k}(t) \big|
\right].
\ee
Note that the intraband current in Eq.~(\ref{eq:intra-current})
is nothing but the current computed solely by
the diagonal elements of the current operator in Eq.~(\ref{eq:current-operator})
in the Houston basis representation.
Because the band velocity of the Dirac bands,
$\partial \epsilon_{b\vec k}/\partial \vec k = \pm v_f \vec k/|\vec k|$,
is isotropic around the Dirac point,
the light-induced Hall current computed 
with Eq.~(\ref{eq:intra-current}) purely originates
from the symmetry breaking of the photocarrier distribution
$n_{b\vec K(t)}(t)$ in the Brillouin zone.

In Fig.~\ref{fig:intensity_scale}, the green squares show
the population imbalance contribution computed by Eq.~(\ref{eq:intra-current}).
In the weak field regime, the population imbalance contribution of
the bare Dirac bands well reproduces the total Hall conductivity (red circles).
Therefore, the light-induced anomalous Hall effect dominantly
originates from the population imbalance of photocarriers.
Furthermore, even in the strong field regime, the population imbalance contribution
fairly captures general trends of the total signal.
Therefore, the population imbalance picture
is still relevant even in the strong field regime.

To obtain further insight into microscopic physics behind
the light-induced anomalous Hall effect, 
we investigate a property of steady states under the presence of circularly-polarized light.
For this purpose, we numerically construct a steady state by
solving Eq.~(\ref{eq:liouville}) under the presence of a continuous circularly-polarized
laser field without the static source-drain field.
After the long time propagation, the system reaches a steady state
due to the balance of the laser excitation and the relaxation.
The single-particle density matrix of such steady state
has a periodicity in time,
$\rho_{\vec K(t)}(t)=\rho_{\vec K(t)}(t+T_{cycle})$, 
with the period of the external field $T_{cycle}$.
To investigate details of such steady state density matrix,
we consider the expansion with natural orbitals \cite{PhysRev.97.1474} as
\be
\rho_{\vec K(t)}(t) = \sum_{b}n^{NO}_{b \vec k}(t)
\big| u^{NO}_{b\vec k}(t)\big\rangle \big \langle u^{NO}_{b\vec k}(t) \big |,
\ee
where $|u^{NO}_{b\vec k}(t)$ is a natural orbital,
and $n^{NO}_{b\vec k}(t)$ is the occupation of the natural orbital.
Thanks to the time periodicity of the density matrix,
the natural orbitals may have the same time periodicity as
$|u^{NO}_{b\vec k}(t+T_{cycle})\rangle=|u^{NO}_{b\vec k}(t)\rangle$.

Importantly, the expectation value of single-particle operators,
Eq.~(\ref{eq:expectation-value}), can be expressed as
a sum of the expectation value of each natural orbital
with the occupation weight as,
\be
\langle \hat O(t) \rangle =  \frac{1}{(2\pi)^2}\int d\vec k
\sum_{b} n^{NO}_{b\vec k}(t)
\big \langle u^{NO}_{b\vec k}(t) \big | \hat O \big |
u^{NO}_{b\vec k}(t) \big \rangle. \nonumber \\
\ee
Therefore, the natural orbital is one of the most suitable descriptors
of the system in a single-particle picture.
Based on this fact, we investigate properties of the steady states
with their natural orbitals.

As seen in Fig.~\ref{fig:intensity_scale}, the contribution from
the population imbalance in the bare Dirac bands~(green-square) 
dominates the total conductivity~(red-circle) in the weak field regime.
Therefore, the steady state in the weak field regime is expected to be well
characterized by the valence and conduction states of the bare Dirac bands.
Once the field strength becomes strong, the contribution from the 
population imbalance mechanism shows a discrepancy from the total conductivity.
Thus, in the strong field regime, the bare valence and conduction states
would not be suitable descriptors of the system anymore.
In order to access this hypothesis, we introduce a measure of similarity
of the natural orbitals and the Houston states~(the instantaneous eigenstates
of the Hamiltonian). We shall call it
\textit{Houston fidelity}.
To define the Houston fidelity, we first introduce a fidelity matrix 
$F^H_{\vec k}$ at each $\vec k$ such that each matrix element $F_{ij,\vec k}$ 
is the cycle-average of the squared overlap of
the $i$-th natural orbital and the $j$-th Houston state as
\be
F_{ij,\vec k} = \frac{1}{T_{cycle}}
\int^{T_{cycle}}_{0} dt \left | 
\big \langle u^{NO}_{i\vec k}(t) \big |
u^{H}_{j\vec k}(t) \big \rangle
\right |^2.
\label{eq:houston-fidelity-matrix}
\ee
Then, the Houston fidelity is defined as the absolute value of 
the determinant of the fidelity matrix,
$S^{H}_{\vec k}=|\mathrm{det}F_{\vec k}|$.
Note that the Houston fidelity, $S^H_{\vec k}$, takes the maximum value of one
only if the natural orbitals are identical to the Houston states at all the time.
In general, $0\le S^{H}_{\vec k}\le 1$.

In contrast to the above bare band picture, 
one may consider a photon-dressed band picture based on
the Floquet theory.
Under the continuous laser driving, the Hamiltonian $H(t)$
has the time periodicity with the period of $T_{cycle}$.
Then, a Floquet state $u^F_{b\vec k}(t)$ may be introduced as
a part of a solution of the time-dependent Schr\"odinger equation
with a time-periodic Hamiltonian,
\be
i\hbar \frac{d}{dt}\big |\Psi^F_{b\vec k}(t)\big\rangle = H_{\vec K(t)}
\big |\Psi^F_{b\vec k}(t)\big\rangle,
\ee
where the solution $|\Psi^F_{b\vec k}(t)\rangle$ 
consists of the time periodic Floquet state, 
$|u^F_{b\vec k}(t+T_{cycle})\rangle=|u^F_{b\vec k}(t)\rangle$,
and a pure phase factor as $|\Psi^F_{b\vec k}(t)\rangle=
\mathrm{exp}[-i\epsilon^F_{b\vec k}t]|u^F_{b\vec k}(t)\rangle$.
Here, $\epsilon^F_{b\vec k}$ is the Floquet quasienergy.
One may further introduce a measure of similarity of the natural orbitals
and the Floquet states by employing the Floquet states in 
Eq.~(\ref{eq:houston-fidelity-matrix}) instead of the Houston states.
This was introduced as $S^F_{\vec k}$ \textit{Floquet fidelity}
in previous work \cite{2019Sato},
and it was used to demonstrate that the Floquet states 
are realized in graphene under the presence of a strong laser field.

\begin{figure}[htbp]
  \centering
  \includegraphics[width=0.6\columnwidth]{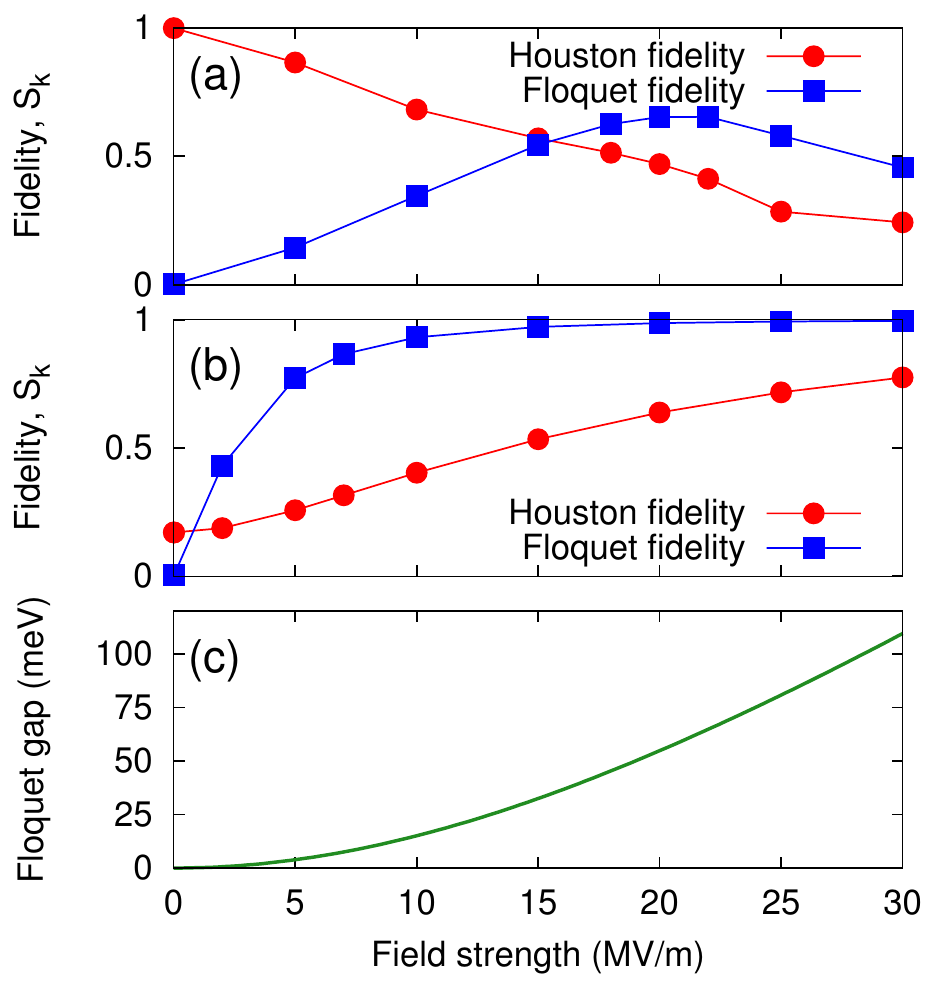}
\caption{\label{fig:fidelity}
Calculated Houston fidelity $S^H_{\vec k}$ and Floquet fidelity $S^F_{\vec k}$
(see main text for details)
as functions of field strength at (a) the resonant $k$-point,
$|k|=\hbar \omega_{cir}/2v_f$ and (b) the Dirac point, $\vec k =0$.
(c) The Floquet gap at the Dirac point as a function of field strength
of the circular laser field.
}
\end{figure}

Since photocarriers are expected to play a significant role
in the light-induced anomalous Hall effect, we now
investigate the steady state density matrix at a resonant $k$-point where
the vertical gap is identical to the photon energy, $|k|=\hbar \omega_{cir}/2v_F$.
Figure~\ref{fig:fidelity}~(a) shows the Houston fidelity $S^H_{\vec k}$
and the Floquet fidelity $S^F_{\vec k}$ at the resonant $k$-point 
as a function of field strength of the circular laser field.
In the weak field regime, the Houston fidelity is close to one,
indicating that the steady state is well characterized by the bare Dirac bands.
As the field strength becomes stronger, the Houston fidelity
monotonically decreases, indicating that the states are significantly dressed
by photons and the system is not well characterized
by the bare Dirac bands anymore. These findings fairly support
the above hypothesis.
The Floquet fidelity $S^F_{\vec k}$ at the resonance in Fig.~\ref{fig:fidelity}~(a)
is close to zero in the weak field regime.
This fact indicates that the photon-dressing electronic state
is significantly disturbed by the dissipation, and the Floquet states
cannot be formed in the weak field regime.
Once the field strength increases, 
the photon-dressing effect becomes more significant
and overcomes the dissipation effects. As a result of the competition
of the dressing and the dissipation, the Floquet states can be fairly well formed
in a strong field regime. However, once the field strength becomes stronger
than $20$~MV/m, the Floquet fidelity starts decreasing.
This fact can be understood by the opening of additional dissipative channels
due to the significant intraband motion of electrons in the Brillouin zone:
Once the field strength becomes strong enough,
the momentum shift of the electron $e|\vec A(t)|/\hbar c$ becomes
comparable to the distance between the Dirac point and the resonant $k$-point
$|k|=\hbar \omega_{cir}/2v_F$. As a result, electrons that originally stay
around the resonant point can move around the Dirac point, where
the gap between the conduction and valence states is small.
Because of the small gap nature, additional dissipative channels can be opened,
and the photon-dressing effect is significantly disturbed.
Indeed, the momentum shift corresponding to the field strength 
of $30$~MV/m is about $8.3\times 10^{-3}$~a.u., and it exceeds
the resonant momentum $|k|=\hbar \omega_{cir}/2v_F=6.8\times 10^{-3}$~a.u.

Next, we investigate the steady state density matrix
at the Dirac point instead of the resonance.
Figure~\ref{fig:fidelity}~(b) shows the Houston fidelity $S^H_{\vec k}$
and the Floquet fidelity $S^F_{b\vec k}$ at the Dirac point, $\vec k=0$.
In contrast to the result at the resonance, the Houston fidelity is
much smaller than one even in the weak field limit.
This fact indicates that the bare band picture is significantly disturbed
by the weak external driving because of the gapless feature at the Dirac point.
The Floquet fidelity at the Dirac point is almost zero in the weak field limit,
while it quickly converges to one as the field strength increases.
Therefore, the Floquet states are well established at the Dirac point
in the strong field regime.
This fact is consistent with the above finding in the resonance condition:
the Floquet states are significantly disturbed by the dissipation in the weak field regime,
while they are stabilized in the strong field regime because the photon-dressing effect
overcomes the dissipation effect.
One sees that the Houston fidelity seems to converge to one in the strong field limit,
indicating that the adiabatic picture becomes suitable in the strong field regime.
However, the realization of the adiabatic states requires much higher field strength,
compared with the Floquet states.

Based on the Floquet picture, it was demonstrated that
the Floquet quasienergy shows the band-gap opening at the Dirac point
under the presence of a circularly-polarized laser field,
and the Floquet states show the emergence of Berry curvatures
\cite{PhysRevB.79.081406}.
Furthermore, the light-induced anomalous Hall effect was predicted
based on the anomalous velocity due to the Floquet Berry curvature.
Figure~\ref{fig:fidelity}~(c) shows the Floquet gap at the Dirac point
as a function of the field strength of the circular laser field.
The gap reaches $100$~meV at the highest field strength, $30$~MV/m.
According to Fig.~\ref{fig:fidelity}~(b), one expects that,
in the strong field regime, the system forms the Floquet states 
and shows the Floquet Berry curvature contribution to the Hall effect
due to the topological gap opening.
Indeed, it was demonstrated that the natural orbitals of 
the non-equilibrium steady state with dissipation
show Berry curvatures consistent with those of the Floquet states
\cite{2019Sato}.
To quantify the Berry curvature contribution to the light-induced anomalous
Hall effect, we extend the expression of the Hall conductivity with
the Floquet states \cite{PhysRevB.79.081406} to that with
the natural orbitals as
\be
\sigma^{B}_{xy}=\frac{2e^2}{\hbar}\int \frac{d\vec k}{(2\pi)^2}
\sum_{b}\tilde n^{NO}_{b\vec k}\tilde \Omega^{NO}_{b\vec k},
\label{eq:signa-berry-curvature}
\ee
where $\tilde n^{NO}_{b\vec k}$ is the cycle-average
natural orbital occupation,
and $\tilde \Omega^{NO}_{b\vec k}$ is the 
cycle-averaged Berry curvature of natural orbitals defined as
\be
\tilde \Omega^{NO}_{b\vec k}=-i\frac{1}{T_{cycle}}
\int^{T_{cycle}}_0dt
\vec \nabla_{\vec k}\times
\big \langle u^{NO}_{b\vec k}(t)\big |\vec \nabla \big |
u^{NO}_{b\vec k}(t) \big \rangle. \nonumber \\
\label{eq:sigma-berry-massless}
\ee

In Fig.~\ref{fig:intensity_scale}, the blue triangles
show the Berry curvature contribution to the light-induced Hall conductivity
computed with Eq.~(\ref{eq:signa-berry-curvature}).
One sees that the light-induced Berry curvature shows non-zero
contribution to the light-induced Hall conductivity.
However, it is a rather minor contribution compared with the population imbalance effect.
Therefore, we can conclude that the light-induced anomalous Hall effect in
graphene dominantly originates from the photocarrier population imbalance
in momentum space.

\subsection{Light-induced Hall effect in topological insulators}

In the previous section, we discussed the emergent Berry curvature contribution
to the light-induced anomalous Hall effect in the massless Dirac fermion system.
Here, we investigate the Hall transport properties of a system
where the light-induced extrinsic contributions coexist with
the intrinsic topological contribution.
For this purpose, we investigate the light-induced anomalous Hall effect
in a massive Dirac fermion system.
We set the gap $\Delta$ to $100$~meV, which is comparable 
to the Flouqet gap of the massless Dirac fermion system at
the highest field strength (see Fig.~\ref{fig:fidelity}~(c)).
The other parameters are the same as the previous massless Dirac fermion system.

In order to investigate the relation between the chirality of 
light~($\tau_{cir}$) and the intrinsic chirality of the system~($\tau_z$),
we do not employ the pump dichroism but employ the cycle-averaged current
of a nonequilibrium steady state under the presence of the source-drain field
and a continuous circular laser field, fixing the chirality of light
to positive~($\tau_{cir}=+1$).

\begin{figure}[htbp]
  \centering
  \includegraphics[width=0.6\columnwidth]{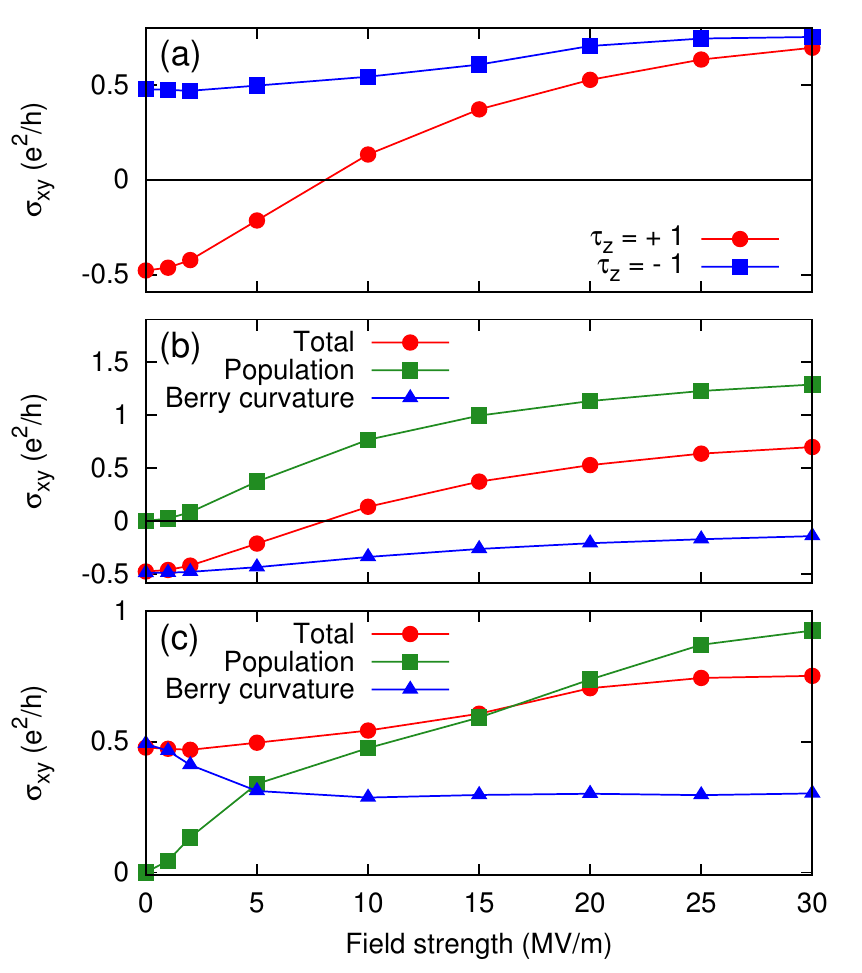}
\caption{\label{fig:field_scaling_massive}
Hall conductivities as a function of the field strength of the applied circular laser fields.
(a) The total Hall conductivity of the massive Dirac fermion system
is shown for the positive chirality system ($\tau_z=+1$) as the red circles
and for the negative chirality system ($\tau_z=-1$) as the blue squares.
(b) The Hall conductivity of the positive chirality system is shown.
The red circles show the total conductivity, the green squares show
the contribution from the population imbalance computed with 
Eq.~(\ref{eq:intra-current}), and the blue triangles show the Berry curvature contribution
computed by Eq.~(\ref{eq:signa-berry-curvature}).
(c) shows now the Hall conductivity for the negative
chirality system.
}
\end{figure}

Figure~\ref{fig:field_scaling_massive}~(a) shows the computed Hall conductivity
$\sigma_{xy}$ as a function of field strength of the circular laser field.
The red circles show the conductivity of the positive chirality 
system~($\tau_z=+1$), while the blue squares show that of the negative chirality
system~($\tau_z=-1$).
In the weak field limit, the two systems show the intrinsic Hall conductivities
with the opposite sign, reflecting the intrinsic Berry curvature contribution.
Once the field strength becomes strong, the two systems with the different
chiralities start showing qualitatively different behaviors
because they have opposite intrinsic contributions and a common extrinsic contribution.
The positive chirality system ($\tau_z=+1$) shows
negative conductivity in the weak field limit. As the field strength increases,
the absolute value of the conductivity decreases.
When the field strength reaches around $8$~MV/m, the Hall conductivity becomes zero
due to the cancellation of the intrinsic and extrinsic contributions.
Once the field strength becomes stronger, the conductivity changes its sign to positive
and monotonically increases.
On the other hand, the negative chirality system ($\tau_z=-1$)
does not show the sign change in the Hall conductivity.
Interestingly, the conductivities of the two systems converge to a similar value
in the strong field regime.
These facts indicate that the common extrinsic contribution to the Hall transport property
overcomes the intrinsic Berry curvature contribution in the strong field regime.

In order to elucidate the microscopic mechanism of the Hall transport property
of the massive Dirac fermion system in the presence of the circular laser field,
we evaluate the contribution from the population imbalance of photocarriers
with Eq.~(\ref{eq:intra-current}) and the Berry curvature contribution
by Eq.~(\ref{eq:signa-berry-curvature}).

Figure~\ref{fig:field_scaling_massive}~(b) shows the result of the positive
chirality system~($\tau_z=+1$).
Here, in the weak field limit, the total conductivity~(red circle)
shows the half integer of the quantized conductivity $-0.5e^2/h$
due to the intrinsic topological contribution, and this value is
well described by the natural-orbital Berry curvature contribution~(blue triangle).
Thus, we can confirm that 
the intrinsic Berry curvature contribution is well captured by
the natural orbital expression, Eq.~(\ref{eq:signa-berry-curvature}).
Furthermore, since there is no photocarrier generation in the weak field limit,
the contribution from the population imbalance~(green square) is zero in this regime.
As the field strength increases, the Berry curvature contribution
becomes smaller while the population imbalance starts being the larger contribution
in the positive chiral system.
As a result, the sign of the Hall conductivity changes from negative to positive.
This fact indicates the robustness of the extrinsic population contribution
and a possibility that the intrinsic property of the system
can be completely overwritten by the extrinsic property
with the strong optical driving.

Figure~\ref{fig:field_scaling_massive}~(c) shows the result of the negative
chirality system~($\tau_z=-1$). 
In contrast to the positive chirality system,
the Hall conductivity of the negative chirality system shows
the positive conductivity in the weak field regime.
Furthermore, the Hall conductivity does not show the sign change
even in the strong field regime.
This behavior can be understood by the fact that, in the negative chirality system,
the intrinsic and the extrinsic contributions have the same sign in the Hall
conductivity (see~Fig.~\ref{fig:field_scaling_massive}~(c)).
In Fig.~\ref{fig:field_scaling_massive}~(c),
while the Berry curvature contribution (blue triangle) dominates 
the intrinsic Hall transport property in the weak field limit,
it is weakened once the field strength becomes strong.
On the other hand, while the contribution from the population imbalance
is negligible in the weak field regime,
it becomes significant in the strong field regime.
Therefore, in both positive and negative chirality systems,
the intrinsic topological contribution is weakened by the laser driving,
while the light-induced extrinsic contribution dominates
the total Hall transport property.

Finally, we investigate the effect of electron (hole) doping on
the anomalous Hall effect in the massive Dirac fermion system.
In previous theoretical and experimental works,
the doping effect on the light-induced anomalous Hall effect in graphene
has been investigated \cite{2018McIver,2019Sato},
and it was demonstrated that
the light-induced Hall conductivity in the strong field regime
shows a double-peak structure with a central plateau as a function of 
the chemical potential.
Here, we elucidate the role of the intrinsic Hall transport property
in the optically-driven massive Dirac fermion system with electron (hole) doping.

\begin{figure}[htbp]
  \centering
  \includegraphics[width=0.6\columnwidth]{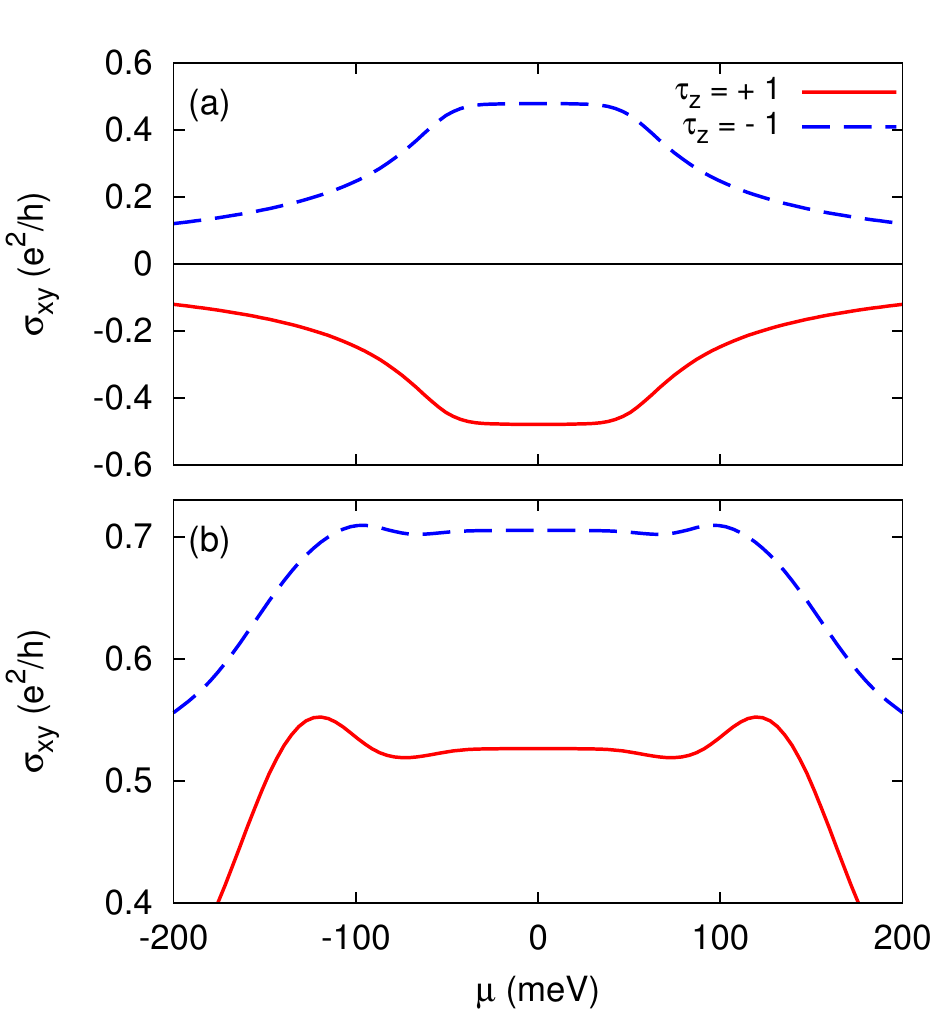}
\caption{\label{fig:sigma_xy_mu_massive}
Hall conductivities of the massive Dirac fermion systems
as a function of chemical potential $\mu$.
(a) The Hall conductivities in equilibrium without optical-driving
are shown for the positive chirality system ($\tau_z=+1$) as the red-solid line
and for the negative chirality system ($\tau_z=-1$) as the blue-dashed line.
(b) The Hall conductivities in a non-equilibrium steady state with
the optical driving~ ($E_{cir}=20$~MV/m) are shown.
}
\end{figure}

Figure~\ref{fig:sigma_xy_mu_massive}~(a) shows the Hall conductivities
of the massive Dirac fermion systems in equilibrium without optical driving
as a function of chemical potential $\mu$,
while Figure~\ref{fig:sigma_xy_mu_massive}~(b) shows those under
the presence of the strong circular laser field~($E_{cir}=20$~MV/m).
In equilibrium, the conductivities show the central plateau region the width of which
is comparable with the size of the gap $\Delta=100$~meV.
Depending on the chirality of the system~($\tau_z = \pm 1$),
the Hall conductivity has the opposite sign.
In contrast, in the nonequilibrium steady state under the presence of the strong
laser field, the Hall conductivities become positive regardless of
the chirality of the system.
Furthermore, both chiral systems show the common feature;
the double peak structure with the central plateau region.
This feature is consistent with the emergent feature of 
the light-induced anomalous Hall effect in graphene.
These findings indicate that, although the systems with opposite chiralities
have significantly different intrinsic properties in equilibrium,
the properties can be overwritten by 
the strong optical drive.

\section{Summary \label{sec:summary}}

In this work, we investigated the transport property of massless
and massive Dirac fermion systems that mimic graphene and
topological insulators, respectively.
Dirac fermion systems under the coexistence of light and dissipation
in order to provide the microscopic insight into the dynamics of optically-driven
nonequilibrium systems.

We assessed two kinds of phenomenological relaxation
operators: One is constructed with the Houston states, which are the instantaneous
eigenstates of Hamiltonian. The other is constructed with the static Bloch states,
which are static eigenstates of the field-free Hamiltonian.
It was demonstrated that the relaxation operator constructed with the Houston states
satisfies Ohm's low for a metallic system and reproduces the quantized Hall conductivity of
a topological insulator while the relaxation operator constructed with
the Bloch states fails to satisfy Ohm's low and to reproduce the quantized conductivity,
showing divergence of the current.
The failure of this relaxation operator can be understood by an artificial excitation
via the relaxation operator because the field-induced intraband motion
is not properly taken into account, which is instead naturally included in the operator
using the Houston states.
Thus, the Houston state expression is indispensable to properly construct 
the dissipation operator and to be able to address the dynamical simulation of
the driven system.

We then investigated the light-induced anomalous Hall effect in a massless
Dirac fermion system that mimics the Dirac bands of graphene.
We demonstrated that the massless Dirac fermion system shows the Hall transport
property under the presence of circular laser fields.
Based on the microscopic analysis, we clarified that the imbalance of
photocarrier distribution in the Brillouin zone is the main origin of
the light-induced Hall transport property although
the nonzero light-induced Berry curvature contribution was also found.
We further analyzed the steady-state density matrix under the continuous circular 
laser driving. As a result, we found that the Floquet states are not well formed
in the weak field regime due to dissipation while 
they are well formed in the strong field regime because the photon-dressing
overcomes the dissipation effect.

Next, we studied the Hall transport property of the massive Dirac fermion system
under the presence of the circular laser fields
in order to investigate the interplay of the intrinsic and the extrinsic contributions
to the Hall transport.
In the weak field regime, the intrinsic topological contribution dominates
the Hall transport property while the photocarrier population effect is negligible.
Once the field strength becomes strong, the topological contribution
becomes weakened while the photocarrier population effect becomes significant.
Surprisingly, if the intrinsic and the extrinsic contributions have opposite signs
to the Hall conductivity, the sing of the Hall conductivity of the topological insulator
can be flipped by the strong laser driving.
Note that, in addition to the intrinsic Berry curvature contribution,
it is known that extrinsic scattering processes such as
the skew-scattering may have a significant contribution to the anomalous Hall effect
in equilibrium phases \cite{RevModPhys.82.1539}. Because the scattering processes
depend much on population distributions of carriers, the photocarrier effect investigated
in this work may further modify the extrinsic scattering contribution and open yet another
possibility of controlling Hall transport properties by light. This possibility will be
investigated by integrating the corresponding scattering mechanisms into the present model
in future work.

The above findings demonstrate the robustness and the generality of
the photocarrier population imbalance effect in the optically-driven nonequilibrium
systems. Importantly, we demonstrated that the extrinsic population effect can overwrite 
intrinsic properties of materials.
Therefore, the population control can be a key to realize the optical control of 
material properties.
Furthermore, by combining the population control with
the state control such as the Floquet engineering by photon-dressing,
properties and functionalities of materials can be largely controlled via light
and novel features in nonequilibrium systems may be discovered.

We acknowledge fruitful discussions with J.W.~McIver, G.~Jotzu, and A.~Cavalleri.
This work was supported by the European Research Council (ERC-2015-AdG694097).
The Flatiron Institute is a division of the Simons Foundation.
S.A.S. gratefully acknowledges the fellowship from the Alexander von Humboldt Foundation.
M.A.S. acknowledges financial support by the DFG through the Emmy Noether programme
(SE 2558/2-1). P.T. acknowledges the received funding from the European Unions 
Horizon 2020 research and innovation programme under the Marie Sklodowska-Curie grant 
agreement No 793609.
A.R. acknowledges support from the Cluster of Excellence 'Advanced Imaging of Matter' (AIM)



\bibliography{ref}

\end{document}